\documentclass[letterpaper]{article} 
\usepackage{aaai24}  
\usepackage{times}  
\usepackage{helvet}  
\usepackage{courier}  
\usepackage[hyphens]{url}  
\usepackage{graphicx} 
\usepackage{natbib}  
\usepackage{caption} 
\usepackage{algorithm}
\usepackage{algorithmic}

\usepackage{newfloat}
\usepackage{listings}
\DeclareCaptionStyle{ruled}{labelfont=normalfont,labelsep=colon,strut=off} 
\floatstyle{ruled}
\newfloat{listing}{tb}{lst}{}
\floatname{listing}{Listing}
\usepackage{drago}
\usepackage[off]{editing}

\title{Reconciling Predictive and Statistical Parity:\\A Causal Approach}
\author{
    Drago Ple\v cko \;{\normalfont and}\;  Elias Bareinboim \\
}
\affiliations{
    Department of Computer Science\\
    Columbia University\\
    New York, NY 10027 \\
    \texttt{dp3144@columbia.edu, eb@cs.columbia.edu} \\
}

\usepackage{selectp}

\begin{document}

\maketitle

\begin{abstract}
Since the rise of fair machine learning as a critical field of inquiry, many different notions on how to quantify and measure discrimination have been proposed in the literature. Some of these notions, however, were shown to be mutually incompatible. Such findings make it appear that numerous different kinds of fairness exist, thereby making a consensus on the appropriate measure of fairness harder to reach, hindering the applications of these tools in practice. In this paper, we investigate one of these key impossibility results that relates the notions of statistical and predictive parity. Specifically, we derive a new causal decomposition formula for the fairness measures associated with predictive parity, and obtain a novel insight into how this criterion is related to statistical parity through the legal doctrines of disparate treatment, disparate impact, and the notion of business necessity. Our results show that through a more careful causal analysis, the notions of statistical and predictive parity are not really mutually exclusive, but complementary and spanning a spectrum of fairness notions through the concept of business necessity. Finally, we demonstrate the importance of our findings on a real-world example.
\end{abstract}

\section{Introduction}
As society increasingly relies on AI-based tools, an ever larger number of decisions that were once made by humans are now delegated to automated systems, and this trend is likely to only accelerate in the coming years. Such automated systems may exhibit discrimination based on gender, race, religion, or other sensitive attributes, as witnessed by various examples in criminal justice \citep{ProPublica}, facial recognition \citep{facedetectionarticle, pmlr-v81-buolamwini18a}, targeted advertising \citep{facebook2019redlining}, and medical treatment allocation \citep{rajkomar2018ensuring}, to name a few.

In light of these challenges, a large amount of effort has been invested in attempts to detect and quantify undesired discrimination based on society's current ethical standards, and then design learning methods capable of removing possible unfairness from future predictions and decisions. During this process, many different notions on how to quantify discrimination have been proposed. In fact, the current literature is abundant with different fairness metrics, some of which are mutually incompatible \citep{corbett2018measure}. The incompatibility of these measures can create a serious obstacle for practitioners since choosing among them, even for the system designer, is usually non-trivial. 

In the real world, issues of discrimination and unfairness are analyzed through two major legal doctrines.
The first one is \textit{disparate treatment}, which enforces the equality of treatment of different groups and prohibits the use of the protected attribute (e.g., race) during the decision process. One of the legal formulations for showing disparate treatment is that ``a similarly situated person who is not a member of the protected class would not have suffered the same fate'' \citep{barocas2016big}. 
Disparate treatment is commonly associated with the notion of direct effects in the causal literature. 
The second doctrine is known as \textit{disparate impact} and focuses on \textit{outcome fairness}, namely, the equality of outcomes among protected groups. Discrimination through disparate impact occurs if a facially neutral practice has an adverse impact on members of the protected group, including cases where discrimination is unintended or implicit. In practice, the law may not necessarily prohibit the usage of all characteristics correlated with the protected attribute due to their relevance to the business itself, which is legally known as ``business necessity'' (labeled BN from now on) or ``job-relatedness''. Therefore, some of the variables may be used to distinguish between individuals, even if they are associated with the protected attribute \citep{kilbertus2017avoiding}. From a causal perspective, disparate impact is realized through indirect forms of discrimination, and taking into account BN considerations is the essence of this doctrine \citep{barocas2016big}. 

\subsection{Relationship to Previous Literature}
Consider now a set of causal pathways between the attribute $X$ and the predictor $\widehat{Y}$, labeled $\mathcal{C}_1, \dots, \mathcal{C}_k$. For example, these pathways could represent the direct, indirect, and spurious effects of $X$ on $\widehat{Y}$, or more generally any set of path-specific effects. Previous works on path-specific counterfactual fairness \citep{nabi2018fair, wu2019pc,chiappa2019path} constrain the effects along paths $\mathcal{C}_i$ that are considered discriminatory (i.e., \textit{not} in the BN set), by requiring them to be equal to 0, written $\mathcal{C}_i(X,\widehat{Y}) = 0$. Other works demonstrate how such causal variations can be disentangled within statistical measures of fairness \citep{zhang2018fairness, zhang2018equality}. 

Importantly, however, these works do not specify anything about causal effects along pathways that \textit{are in the BN set}, i.e., are not considered discriminatory. If $\mathcal{C}_i$ is in the BN set, the transmitted causal effect does not need to equal $0$ in general, i.e., $\mathcal{C}(X, \widehat{Y}) \neq 0$ may be allowed. In this paper, we demonstrate that a natural way of constraining effects along BN paths exists, and the transmitted causal effect should not take an arbitrary value in this case. Rather, it should equal the \textit{transmitted effect from $X$ to the original outcome $Y$ (observed in the real world) along the same pathway}, written as $\mathcal{C}_i(X, \widehat{Y}) = \mathcal{C}_i(X, Y)$. In this way, we complement the existing literature on path-specific fairness.

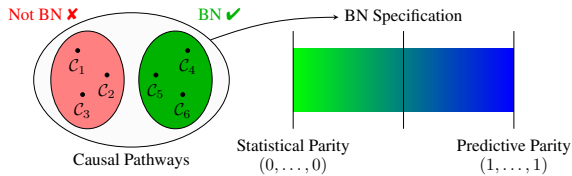
\begin{figure}
\centering
    \vspace{-0.05in}
    \scalebox{0.65}{
    \begin{tikzpicture}[>=stealth]
  \node[ellipse, draw, fill=gray!3, minimum width=4cm, minimum height=2.75cm, label=below:Causal Pathways] (ellipse) at (-3.5,0) {};
  \node[ellipse, draw, thin, fill=red!50, minimum width=1.5cm, minimum height=2cm, label={[red, xshift=-0.9cm, yshift=0.1cm]Not BN \mbox{\ding{56}}}] (ellipse1) at (-4.4,0) {};
  \node[ellipse, draw, thin, fill=black!30!green, minimum width=1.5cm, minimum height=2cm, label={[black!30!green, xshift=0.9cm, yshift=0.1cm]BN \mbox{\ding{52}}}] (ellipse2) at (-2.6,0) {};
  \node[circle, fill=black, inner sep=1pt, label=below:$\mathcal{C}_1$] (point1) at (-4.6,0.6) {};
  \node[circle, fill=black, inner sep=1pt, label=below:$\mathcal{C}_2$] (point1) at (-4.0,0.1) {};
  \node[circle, fill=black, inner sep=1pt, label=below:$\mathcal{C}_3$] (point1) at (-4.5,-0.3) {};
  \node[circle, fill=black, inner sep=1pt, label=below:$\mathcal{C}_4$] (point2) at (-2.35,0.6) {};
  \node[circle, fill=black, inner sep=1pt, label=below:$\mathcal{C}_5$] (point2) at (-3,0.1) {};
  \node[circle, fill=black, inner sep=1pt, label=below:$\mathcal{C}_6$] (point2) at (-2.45,-0.3) {};
  \node[rectangle, right=1.3cm of ellipse, minimum width=4.5cm, minimum height=1.3cm, shading=axis, left color=green, right color=blue] (rectangle) {};
  \draw (rectangle.west) ++ (0,-1) node[below, align=center,yshift=-0.1cm] {Statistical Parity\\ $(0, \dots, 0)$} -- ++ (0,2) ;
  \draw (rectangle.east) ++ (0, -1) node[below, align=center,yshift=-0.1cm] {Predictive Parity\\ $(1, \dots, 1)$} -- ++ (0,2);
  \draw (rectangle.center) ++ (0,-1) -- ++ (0,2) node[above] (bn-spec) {BN Specification};
  \draw[->] (ellipse) edge[bend left=10] (bn-spec.west);
\end{tikzpicture}
    }
    \caption{BN specification (left) and the spectrum between statistical and predictive parity (right).}
    \label{fig:sp-pp-spectrum}
    \vspace{-0.15in}
\end{figure}

As we will demonstrate, considerations of business necessity are also closely related to the impossibility result between statistical and predictive parity \citep{barocas2017fairness}. When reasoning about fairness, the system designer needs to decide which of the causal pathways $\mathcal{C}_1, \dots, \mathcal{C}_k$ are considered discriminatory, and which ones are not, as shown in Fig.~\ref{fig:sp-pp-spectrum} (left side). Therefore, considerations of BN can be summarized as a $0/1$ vector of length $k$, where each pathway $\mathcal{C}_i$ is represented by an entry.
As we demonstrate both intuitively and formally throughout the paper, the choice of the BN set being empty, written $(0, \dots, 0)$, will ensure the notion of statistical parity. 
The choice of the BN set that includes all pathways, written $(1, \dots, 1)$, will lead to predictive parity. 
Crucially, various intermediate fairness notions between the two ends of the spectrum are possible, depending on what is allowed or not; for an illustration, see Fig.~\ref{fig:sp-pp-spectrum} (right side). 

The unification of the principles behind statistical and predictive parity through the concept of BN offers a new perspective, by providing an argument against the seemingly discouraging impossibility result between them. The practitioner is no longer faced with a false dichotomy of choosing between statistical or predictive parity but rather faces a spectrum of different fairness notions determined by the choice of the business necessity set, which is usually fixed through societal consensus and legal requirements.

Finally, we also mention the work on counterfactual predictive parity (Ctf-PP) \citep{coston2020counterfactual} that is similar in name to our notion of causal predictive parity, but is in fact a very different notion. Ctf-PP deals with the setting of decision-making and considers counterfactuals of the outcome $Y$ with respect to a treatment decision $D$ that precedes it, while our work considers counterfactuals of the outcome $Y$ and the predictor $\widehat{Y}$ with respect to the protected attribute $X$, in the context of fair predictions, and thus offers a rather different line of reasoning.

\subsection{Organization \& Contributions}
In Sec.~\ref{sec:background}, we introduce important preliminary notions in causal inference, and the formal tools for specifying causal pathways $\mathcal{C}_1, \dots, \mathcal{C}_k$ described above. Further, we discuss the notions of statistical and predictive parity, together with the impossibility result that separates them.   
In Sec.~\ref{sec:tv-decomposition} we discuss how different causal variations with the statistical parity measure can be disentangled using an additive decomposition based on the previous result of \cite{zhang2018fairness}. 
In Sec.~\ref{sec:causal-decompositions}, we develop a novel decomposition of the predictive parity measure in terms of the underlying causal mechanisms and discuss how this notion is in fact complementary to statistical parity through the concept of business necessity. In Sec.~\ref{sec:combining-sp-pp}, we unify the theoretical findings by introducing a formal procedure that shows how to assess the legal doctrines of discrimination by leveraging both the concepts of causal predictive parity and causal statistical parity. In Sec.~\ref{sec:experiment}, we apply our approach in the context of criminal justice using the COMPAS dataset \citep{ProPublica}, and demonstrate empirically the trade-off between SP and PP. Our key formal contributions are the following:
\begin{itemize}
    \item We develop the first non-parametric decomposition of the predictive parity measure in terms of the underlying causal mechanisms (Thm.~\ref{thm:pp-decomposition}). 
     \item Building on the previous result, we define a natural notion of causal predictive parity (Def.~\ref{def:causal-pp}).  We then develop a procedure (Alg.~\ref{algo:sp-to-pp-bn}) for evaluating if a classifier satisfies the desired notions of causal statistical parity and causal predictive parity. This provides a unified framework for incorporating considerations of business necessity and sheds light on the impossibility theorem between statistical and predictive parity.
\end{itemize}
\section{Background} \label{sec:background}
We use the language of structural causal models (SCMs) as our basic semantical framework \citep{pearl:2k}. A structural causal model (SCM) is
a tuple $\mathcal{M} := \langle V, U, \mathcal{F}, P(u)\rangle$ , where $V$, $U$ are sets of
endogenous (observables) and exogenous (latent) variables, 
respectively, $\mathcal{F}$ is a set of functions $f_{V_i}$,
one for each $V_i \in V$, where $V_i \gets f_{V_i}(\pa(V_i), U_{V_i})$ for some $\pa(V_i)\subseteq V$ and
$U_{V_i} \subseteq U$. $P(u)$ is a strictly positive probability measure over $U$. Each SCM $\mathcal{M}$ is associated to a causal diagram $\mathcal{G}$ \citep{pearl:2k} over the node set $V$ where $V_i \rightarrow V_j$ if $V_i$ is an argument of $f_{V_j}$, and $V_i \bidir V_j$ if the corresponding $U_{V_i}, U_{V_j}$ are not independent. An instantiation of the exogenous variables $U = u$ is called a \textit{unit}. By $Y_{x}(u)$ we denote the potential response of $Y$ when setting $X=x$ for the unit $u$, which is the solution for $Y(u)$ to the set of equations obtained by evaluating the unit $u$ in the submodel $\mathcal{M}_x$, in which all equations in $\mathcal{F}$ associated with $X$ are replaced by $X = x$.
Building on the notion of a potential response, one can further define the notions of counterfactual and factual contrasts, given by:
\begin{definition}[Contrasts \citep{plecko2022causal}] \label{def:contrast}
Given an SCM $\mathcal{M}$,  a contrast $\mathcal{C}$ is any quantity of the form 
\begin{equation} \label{eq:contrast}
    \contrast = \ex[y_{C_1} \mid E_1] - \ex[y_{C_0}\mid E_0],
\end{equation}
where $E_0, E_1$ are observed (factual) events and $C_0, C_1$ are counterfactual clauses to which the outcome $Y$ responds. A clause $C$ is an intervention of the form $V' = v'$ where $V'$ is a subset of the observable $V$. Furthermore, whenever 
\begin{enumerate}[label=(\alph*)]
    \item $E_0 = E_1$, the contrast $\mathcal{C}$ is said to be counterfactual; 
    \item $C_0 = C_1$, the contrast  $\mathcal{C}$ is said to be factual.
\end{enumerate}
\end{definition}
For instance, the contrast $(C_0 = \{x_0\}, C_1 = \{x_1\}, E_0 = \emptyset, E_1=\emptyset)$ corresponds to the \textit{average treatment effect (ATE)} $\ex[y_{x_1}-y_{x_0}]$. Similarly, the contrast $(C_0 = \{x_0\}, C_1 = \{x_1\}, E_0 = \{x_0\}, E_1=\{x_0\})$ corresponds to the \textit{effect of treatment on the treated (ETT)} $\ex[y_{x_1}-y_{x_0} \mid x_0]$. Many other important causal quantities can be represented as contrasts, as exemplified later on.

\begin{figure}
    \centering
        \centering
	\scalebox{1}{
        \begin{tikzpicture}
	 [>=stealth, rv/.style={thick}, rvc/.style={triangle, draw, thick, minimum size=7mm}, node distance=18mm]
	 \pgfsetarrows{latex-latex};
	 \begin{scope}
		\node[rv] (0) at (0,0.8) {$Z$};
	 	\node[rv] (1) at (-1.5,0) {$X$};
	 	\node[rv] (2) at (0,-0.8) {$W$};
	 	\node[rv] (3) at (1.5,0.6) {$Y$};
            \node[rv] (4) at (1.5,-0.6) {$\widehat{Y}$};
	 	\draw[->] (1) -- (2);
		\draw[->] (0) -- (3);
            \draw[->] (0) -- (4);
	 	\path[->] (1) edge[bend left = 0] (3);
            \path[->] (1) edge[bend left = 0] (4);
		\path[<->] (1) edge[bend left = 30, dashed] (0);
	 	\draw[->] (2) -- (3);
            \draw[->] (2) -- (4);
		\draw[->] (0) -- (2);
	 \end{scope}
	 \end{tikzpicture}
        }
    \caption{Standard Fairness Model.}
    \label{fig:sfm}
\end{figure}
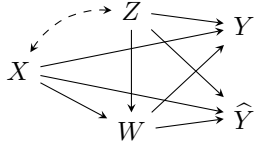

Throughout this manuscript, we assume a specific cluster causal diagram $\mathcal{G}_{\text{SFM}}$ known as the standard fairness model (SFM) \citep{plecko2022causal} over endogenous variables $\{X, Z, W, Y, \widehat{Y}\}$ shown in Fig.~\ref{fig:sfm}. The SFM consists of the following: \textit{protected attribute}, labeled $X$ (e.g., gender, race, religion), assumed to be binary; the set of \textit{confounding} variables $Z$, which are not causally influenced by the attribute $X$ (e.g., demographic information, zip code); the set of \textit{mediator} variables $W$ that are possibly causally influenced by the attribute (e.g., educational level or other job-related information); the \textit{outcome} variable $Y$ (e.g., GPA, salary); the \textit{predictor} of the outcome $\widehat{Y}$ (e.g., predicted GPA, predicted salary). The SFM also encodes the assumptions typically used in the causal inference literature about the lack of hidden confounding\footnote{Partial identification (bounding) of effects can be used to relax these assumptions \citep{zhang2022partialctf}.}.
We next introduce the key notions and results from the fair ML literature needed for our discussion.
\subsection{Statistical \& Predictive Parity Notions} \label{sec:sp-pp-prelim}
The notions of statistical parity (labeled SP from now on) and predictive parity (labeled PP) are defined as follows:
\begin{definition}[Statistical \citep{darlington1971} and Predictive \citep{chouldechova2017fair} Parity] \label{def:stat-parity}
Let $X$ be the protected attribute, $Y$ the true outcome, and $\widehat{Y}$ the outcome predictor. The predictor $\widehat{Y}$ satisfies statistical parity (SP) with respect to $X$ if $X \ci \widehat{Y}$, or equivalently if the statistical parity measure (SPM, for short) satisfies:
\begin{align}
    \text{SPM}_{x_0, x_1}(\widehat{y}) := P(\widehat{y} \mid x_1) - P(\widehat{y} \mid x_0) = 0.
\end{align}
Further, $\widehat{Y}$ satisfies predictive parity (PP) with respect to $X, Y$ if $Y \ci X \mid \widehat{Y}$, or equivalently if $\;\forall \widehat{y}$ we have
\begin{align}
    \text{PPM}_{x_0, x_1}(y\mid \widehat{y}) := P(y \mid x_1, \widehat{y}) - P(y \mid x_0, \widehat{y}) = 0 . \label{eq:pp-measure}
\end{align}
\end{definition}
Statistical parity ($\widehat{Y} \ci X$) requires that the predictor $\widehat{Y}$ contains no information about the protected attribute $X$. In contrast to this, the notion of predictive parity ($Y \ci X \mid \widehat{Y}$), requires that $\widehat{Y}$ should ``exhaust'', or capture all the variations of $X$ coming into the outcome $Y$ in the current real world. \citet{chouldechova2017fair} introduced predictive parity for a binary predictor $\widehat{Y}$, whereas the condition was termed calibration for a continuous classification score. In this paper, we are agnostic to this distinction, and almost all of the results are unaffected by this difference, as discussed in detail in Appendix~\ref{appendix:binary-vs-cts}. Therefore, we simply use the term predictive parity throughout.
An important known result on predictive parity is the following (see proof in Appendix~\ref{appendix:prop1-proof}):
\begin{proposition}[PP and Efficient Learning] \label{prop:pp-efficient}
Let $\mathcal{M}$ be an SCM compatible with the Standard Fairness Model (SFM). Suppose that the predictor $\widehat{Y}$ is based on the features $X, Z, W$. Suppose also that $\widehat{Y}$ is an efficient learner, i.e.:
\begin{align}
    \widehat{Y}(x, z, w) = P(y \mid x, z, w).\label{eq:efficient-learner}
\end{align}
Then $\widehat{Y}$ satisfies predictive parity w.r.t. $X$ and $Y$.
\end{proposition}
Prop.~\ref{prop:pp-efficient} that pertains to a continuous score-based predictor shows that PP is expected to hold whenever we learn the true conditional distribution $P(y \mid x, z, w)$. 
\begin{figure}
        \centering
    \begin{tikzpicture}
	 [>=stealth, rv/.style={thick}, rvc/.style={triangle, draw, thick, minimum size=7mm}, node distance=18mm]
	 \pgfsetarrows{latex-latex};
	 \begin{scope}
	 	\node[rv] (x) at (-1.5,-1) {$X$};
	 	\node[rv] (w) at (0,-2.2) {$W$};
	 	\node[rv] (y) at (1.5,-1) {$Y$};
	 	\node[rv] (yhat) at (1.5,-2.2) {$\widehat{Y}$};
	 	\draw[->] (x) -- (w);
	 	\draw[->] (x) -- (y);
		\draw[->] (x) -- (yhat);
		\draw[->] (w) -- (yhat);
		\draw[->] (w) -- (y);
	 \end{scope}
	 \end{tikzpicture}
    \caption{Standard Fairness Model with $Z = \emptyset$ from Thm.~\ref{thm:pp-decomposition}, extended with the predictor $\widehat{Y}$.}
    \label{fig:pp-decomposition-graph}
\end{figure}
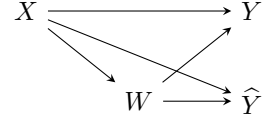
This is in stark contrast with the SP notion. For the SP to hold, the predictor $\widehat{Y}$ should not be associated at all with the attribute $X$, whereas for PP to hold, $\widehat{Y}$ should contain all variations coming from $X$. Perhaps unsurprisingly after this discussion, the following impossibility result holds:
\begin{proposition}[SP and PP impossibility \citep{barocas2017fairness}] \label{prop:dp-pp-impossibility}
Statistical parity ($\widehat{Y} \ci X$) and predictive parity ($Y \ci X \mid \widehat{Y}$) are mutually exclusive except in degenerate cases when $Y \ci X$. 
\end{proposition}
The above theorem might lead the reader to believe that SP and PP criteria come from two different realities, and bear no relation to each other. After all, the theorem states that it is not possible for the predictor $\widehat{Y}$ to include \textit{all variations} of $X$ coming into $Y$ and simultaneously include \textit{no variations} of $X$ coming into $Y$. This realization is the starting point of our discussion in the rest of the manuscript.

\subsection{Statistical Parity Decomposition} \label{sec:tv-decomposition}
We continue by analyzing the decomposition of the statistical parity measure (also known as parity gap, or total variation), which is commonly used to determine if statistical parity is satisfied. The causal decomposition of the SPM requires the usage of causal measures known as counterfactual direct, indirect, and spurious effects, which are defined as:
\begin{definition}[Counterfactual Causal Measures]
   The counterfactual-\{direct, indirect, spurious\} effects of $X$ on $\widehat{Y}$ are defined as follows 
    \begin{align}
\text{Ctf-DE}_{x_0, x_1}(\widehat{y}\mid x) &= P(\widehat{y}_{x_1, W_{x_0}} \mid x) - P(\widehat{y}_{x_0}\mid x) \\
\text{Ctf-IE}_{x_1, x_0}(\widehat{y}\mid x) &= P(\widehat{y}_{x_1, W_{x_0}}\mid x) - P(\widehat{y}_{x_1} \mid x) \\
\text{Ctf-SE}_{x_1, x_0}(\widehat{y}) &= P(\widehat{y}_{x_1} \mid x_0) - P(\widehat{y}_{x_1} \mid x_1).
    \end{align}
\end{definition}
The measures capture the variations from $X$ to $\widehat{Y}$ going through: (i) the direct mechanism $X \rightarrow \widehat{Y}$; (ii) the indirect mechanism $X \rightarrow W \rightarrow \widehat{Y}$; (iii) the spurious mechanisms $X \bidir Z \rightarrow \widehat{Y}$, and $X \bidir Z \rightarrow W \rightarrow \widehat{Y}$, respectively (see Fig.~\ref{fig:sfm}). 
Based on the defined measures, SPM$_{x_0,x_1}(\widehat{y})$ admits an additive decomposition, first obtained in \citep{zhang2018fairness}, given in the following theorem\footnote{We note that the same decomposition can also be applied for the true outcome $Y$ instead of the predictor $\widehat{Y}$, as will be relevant in later sections.}:
\begin{proposition}[Causal Decomposition of Statistical Parity \citep{zhang2018fairness}] \label{prop:tv-decomp}
The statistical parity measure admits a decomposition into its direct, indirect, and spurious variations:
    \begin{align}
        \text{SPM}_{x_0, x_1}(\widehat{y}) 
        =&\; {\text{Ctf-DE}_{x_0, x_1}(\widehat{y}\mid x_0)} - {\text{Ctf-IE}_{x_1, x_0}(\widehat{y}\mid x_0)} \nonumber \\ &- {\text{Ctf-SE}_{x_1, x_0}(\widehat{y})}.
    \end{align}
\end{proposition}
This result shows how we can disentangle direct, indirect, and spurious variations within the SPM. We emphasize the importance of this result in the context of assessing the legal doctrines of discrimination. If a causal pathway (direct, indirect, or spurious) does not lie in the business necessity set, then the corresponding counterfactual measure (Ctf-DE, IE, or SE) needs to equal $0$. This notion follows directly from the literature on path-specific fairness \citep{nabi2018fair, chiappa2019path, wu2019pc}, and we refer to it as \textit{causal statistical parity}:
\begin{definition}[Causal Statistical Parity] \label{def:causal-sp}
We say that $\widehat{Y}$ satisfies causal statistical parity with respect to the protected attribute $X$ if
\begin{align}
    {\text{Ctf-DE}_{x_0, x_1}(\widehat{y}\mid x_0)} &= {\text{Ctf-IE}_{x_1, x_0}(\widehat{y}\mid x_0)} \\
    &= {\text{Ctf-SE}_{x_1, x_0}(\widehat{y})} = 0.
\end{align}
\end{definition}
In practice, causal statistical parity can be a strong requirement, but the notion can be easily relaxed to include only a subset of the Ctf-\{DE, IE, or SE\} measures, under BN requirements.
\section{Predictive Parity Decomposition} \label{sec:causal-decompositions}
After discussing the decomposition of the SPM, our aim is to obtain a causal understanding of the predictive parity criterion, $Y \ci X \mid \widehat{Y}$. To do so, we derive a formal decomposition result of the PP measure that involves both $Y$ and $\widehat{Y}$, shown in the following theorem:
		
\begin{theorem}[Causal Decomposition of Predictive Parity]\label{thm:pp-decomposition}
Let $\mathcal{M}$ be an SCM compatible with the causal graph in Fig.~\ref{fig:pp-decomposition-graph} (i.e., SFM with $Z=\emptyset$). Then, it follows that the PPM$_{x_0,x_1}(y \mid \widehat{y})=P(y \mid x_1, \widehat{y}) - P(y \mid x_0, \widehat{y})$ can be decomposed into its causal and spurious anti-causal variations as:
\begin{align}\label{eq:pp-decomp-thm}
     \text{PPM}_{x_0,x_1}(y \mid \widehat{y})=& \;P(y_{x_1} \mid x_1, \widehat{y}) - P(y_{x_0} \mid x_1, \widehat{y})
    \\&+ P(y_{x_0} \mid \widehat{y}_{x_1}) - P(y_{x_0} \mid \widehat{y}_{x_0}). \nonumber
\end{align}
\end{theorem}
Thm.~\ref{thm:pp-decomposition} offers a non-parametric decomposition result of the predictive parity measure that can be applied to any SCM compatible with the graph in Fig.~\ref{fig:pp-decomposition-graph}. In Appendix~\ref{appendix:pp-decomp-proof} we provide a proof of the theorem (together with the proof of Cor.~\ref{cor:linear-pp-decomp} stated below), and in Appendix~\ref{appendix:pp-decomp-empirical} we perform an empirical study to verify the decomposition result of the theorem. For the additional special case of linear SCMs, the terms appearing in the decomposition in Eq.~\ref{eq:pp-decomp-thm} can be computed explicitly:
\begin{corollary}[Causal Decomposition of Predictive Parity in the Linear Case] \label{cor:linear-pp-decomp}
Under the additional assumptions that (i) the SCM $\mathcal{M}$ is linear and $Y$ is continuous; (ii) the learner $\widehat{Y}$ is efficient, then 
\begin{align}
    \ex(y_{x_1} \mid x_1, \widehat{y}) - \ex(y_{x_0} \mid x_1, \widehat{y}) &= \alpha_{XW}\alpha_{WY} + \alpha_{XY} \label{eq:pp-causal} \\
    \ex(y_{x_0} \mid x_1, \widehat{y}_{x_1}) - \ex(y_{x_0} \mid x_1, \widehat{y}_{x_0}) &= -(\alpha_{XW}\alpha_{WY} + \alpha_{XY}),
\end{align}
where $\alpha_{V_iV_j}$ is the linear coefficient between variables $V_i, V_j$.
\end{corollary}
We now carefully unpack the key insight from Thm.~\ref{thm:pp-decomposition}. In particular, we showed that in the case of an SFM with $Z = \emptyset$\footnote{We remark that the essence of the argument is unchanged in the case with $Z \neq \emptyset$, but handling this case limits the clarity of presentation.} the predictive parity measure can be written as:
\begin{align} \label{eq:pp-decomposition}
    \text{PPM} = &\underbrace{P(y_{x_1} \mid x_1, \widehat{y}) - P(y_{x_0} \mid x_1, \widehat{y})}_{\text{Term (I) causal}} \\ &+ \underbrace{P(y_{x_0} \mid \widehat{y}_{x_1}) - P(y_{x_0} \mid \widehat{y}_{x_0})}_{\text{Term (II) reverse-causal spurious}}. \nonumber 
\end{align}
Term (I) capturing causal variations can be expanded as
\begin{align}
    \sum_{u} \big[\underbrace{y_{x_1}(u) - y_{x_0}(u)}_{\text{unit-level difference}}\big] \underbrace{P(u \mid x_1, \widehat{y})}_{\text{posterior}}.
\end{align}
The expansion shows us that the term is a weighted average of unit-level differences $y_{x_1}(u) - y_{x_0}(u)$. Each unit-level difference measures the causal effect of a transition $x_0 \to x_1$ on $Y$. The associated weights are given by the posterior $P(u \mid x_1, \widehat{y})$, which determines the probability mass corresponding to a unit $u$ within the set of all units compatible with $x_1, \widehat{y}$. Therefore, the term represents an average causal effect of $X$ on $Y$ for a specific group of units chosen by the $x_1, \widehat{y}$ conditioning. Interestingly, after selecting the units by $x_1, \widehat{y}$, the effect in Term (I) \textit{no longer depends on the constructed predictor} $\widehat{Y}$, but only on the true mechanism of $Y$ that already exists in the real world, i.e., it is not under the control of the predictor designer.
The additional linear result in Cor.~\ref{cor:linear-pp-decomp} may also help the reader ground this idea, since it shows that Term (I) indeed captures the causal effect, which, in the linear case, can be obtained using the path-analysis of \citep{wright1934method}.

Thus, to achieve the criterion $\text{PPM} = 0$, the second term needs to be exactly the reverse of the causal effect, captured by the spurious variations induced by changing $\widehat{y}_{x_0} \to \widehat{y}_{x_1}$ in the selection of units. The second term, which is in the control of the predictor $\widehat{Y}$ designer, needs to cancel out the causal effect measured by the first term for PPM to vanish. Therefore, we see that achieving predictive parity is about constructing $\widehat{Y}$ in a way that reflects the causal effect of $X$ on $Y$, across various groups of units. This key observation motivates a novel definition of \textit{causal predictive parity}:
\begin{definition}[Causal Predictive Parity] \label{def:causal-pp}
Let $\widehat{Y}$ be a predictor of the outcome $Y$, and let $X$ be the protected attribute. 
Then, $\widehat{Y}$ is said to satisfy causal predictive parity (CPP) with respect to a counterfactual contrast $(C_0, C_1, E, E)$ if
\begin{align}
     \ex[y_{C_1} \mid E] - \ex[y_{C_0} \mid E] = \ex[\widehat{y}_{C_1} \mid E] - \ex[\widehat{y}_{C_0} \mid E].
\end{align}
Furthermore, $\widehat{Y}$ is said to satisfy CPP with respect to a factual contrast $(C, C, E_0, E_1)$ if
\begin{align}
    \ex[y_{C} \mid E_1] - \ex[y_{C} \mid E_0] = \ex[\widehat{y}_{C} \mid E_1] - \ex[\widehat{y}_{C} \mid E_0].
\end{align}
\end{definition}
The intuition behind the notion of causal predictive parity captures the intuition behind predictive parity. If a contrast $\mathcal{C}$ describes some amount of variation in the outcome $Y$, then it should describe the same amount of variation in the predicted outcome $\widehat{Y}$. For any of the contrasts Ctf-\{DE, IE, SE\} corresponding to a causal pathway, causal predictive parity would require that $\mathcal{C}(X, \widehat{Y}) = \mathcal{C}(X, Y)$.

\subsection{Reconciling Statistical and Predictive Parity} \label{sec:combining-sp-pp}
\begin{algorithm}[tb] 
   \caption{Business Necessity Cookbook}
\begin{algorithmic}[1]
   \STATE {\bfseries Input:} data $\mathcal{D}$, BN-Set $\subseteq \{ \text{DE}, \text{IE}, \text{SE}\}$, predictor $\widehat{Y}$
   \FOR{CE $\in \{ \text{DE}, \text{IE}, \text{SE}\}$}
    \IF{CE $\in$ BN}
        \STATE Compute the effects $\text{Ctf-CE}(y)$, $\text{Ctf-CE}(\widehat{y})$
        \STATE Assert that $\text{Ctf-CE}(y) = \text{Ctf-CE}(\widehat{y})$, otherwise FAIL
    \ELSE
        \STATE Compute the effect $\text{Ctf-CE}(\widehat{y})$
        \STATE Assert that $\text{Ctf-CE}(\widehat{y}) =  0$, otherwise FAIL
    \ENDIF
   \ENDFOR
   \IF{not FAIL} \STATE return SUCCESS \ENDIF
   \STATE {\bfseries Output:} SUCCESS or FAIL of ensuring that disparate impact and treatment hold under BN.
\end{algorithmic}
    \label{algo:sp-to-pp-bn}
\end{algorithm}
We now tie the notions of statistical and predictive parity through the concept of \textit{business necessity}. In particular, if a contrast $\mathcal{C}$ is associated with variations that are not in the business necessity set, then the value of this contrast should be $\mathcal{C}(X, \widehat{Y}) = 0$, following the intuition of causal statistical parity from Def.~\ref{def:causal-sp}. However, if the variations associated with the contrast \textit{are} in the business necessity set, then the value of that contrast should be equal for the predictor to the value for the true outcome
\begin{align}
    \mathcal{C}(X, \widehat{Y}) = \mathcal{C}(X, Y), \label{eq:causal-pp-contrast-2nd}
\end{align}
following the intuition of causal predictive parity. Combining these two notions through business necessity results in Alg.~\ref{algo:sp-to-pp-bn}. The algorithm requires the user to compute the measures
\begin{align}
    \text{Ctf-\{DE, IE, SE\}}(y), \text{Ctf-\{DE, IE, SE\}}(\widehat{y}). 
\end{align}
Importantly, under the SFM, these measures are \textit{identifiable} from observational data:
\begin{proposition} \label{prop:ctf-family-id}
Under the assumptions of the standard fairness model in Fig.~\ref{fig:sfm}, the causal measures of fairness $\text{Ctf-\{DE, IE, SE\}}(y), \text{Ctf-\{DE, IE, SE\}}(\widehat{y})$ are identifiable from observational data, that is, they can be computed uniquely from the observational distribution $P(V)$. 
\end{proposition}
The explicit identification expression for each of the measures is given in Appendix~\ref{appendix:measure-id-expressions}. The above result guarantees that the procedure of Alg.~\ref{algo:sp-to-pp-bn} is applicable to practical data analysis, in a fully non-parametric nature.

Further, for each of the DE, IE, and SE effects, the user needs to determine whether the causal effect (CE) in question falls into the business necessity set. If yes, then the algorithm asserts that
\begin{align}
    \text{Ctf-CE}(y) = \text{Ctf-CE}(\widehat{y}).
\end{align}
In the other case, when the causal effect is not in the business necessity set, the algorithm asserts that
\begin{align}
    \text{Ctf-CE}(\widehat{y}) = 0.
\end{align}
Alg.~\ref{algo:sp-to-pp-bn} is the key result of this paper, and provides a practical approach to incorporating ideas of business necessity through a causal lens. We also remark that it is written in its population level version, in which the causal effects are estimated perfectly with no uncertainty. In the finite sample case, one needs to perform hypothesis testing to see whether the effects differ. If one is also interested in constructing a new fair predictor before using Alg.~\ref{algo:sp-to-pp-bn} (instead of testing an existing one), one may use tools for causally removing discrimination, such as \citep{chiappa2019path} or \citep{plevcko2020fair, plevcko2021fairadapt}. In Appendix~\ref{appendix:algo-1-existence} we show formally that a predictor $\widehat Y$ satisfying the conditions of Alg.~\ref{algo:sp-to-pp-bn} can be constructed for any choice of the BN-set.

Finally, we also remark that Alg.~\ref{algo:sp-to-pp-bn} generalizes to arbitrary causal diagrams $\mathcal{G}$. For the SFM, we can only distinguish three causally different pathways (direct, indirect, spurious), whereas for arbitrary graphs we may consider various collections of path-specific effects. If the contrast $\mathcal{C}_\pi$ captures variations along a pathway $\pi$, then Alg.~\ref{algo:sp-to-pp-bn} would simply test the equality $\mathcal{C}_\pi(X, \widehat{Y}) = \mathcal{C}_\pi(X, Y)$ if $\pi$ is in the BN set. For larger graphs, however, the number of BN paths may scale unfavorably, representing a computational challenge.

\section{Experiments} \label{sec:experiment}
We now apply Alg.~\ref{algo:sp-to-pp-bn} to the COMPAS dataset \citep{ProPublica}, as described in the following example. Additionally, in Appendix~\ref{appendix:census}, we also analyze the Census 2018 dataset.

Courts in Broward County, Florida use machine learning algorithms, developed by Northpointe, to predict whether individuals released on parole are at high risk of re-offending within 2 years ($Y$). The algorithm is based on the demographic information $Z$ ($Z_1$ for gender, $Z_2$ for age), race $X$ ($x_0$ denoting White, $x_1$ Non-White), juvenile offense counts $J$, prior offense count $P$, and degree of charge $D$.

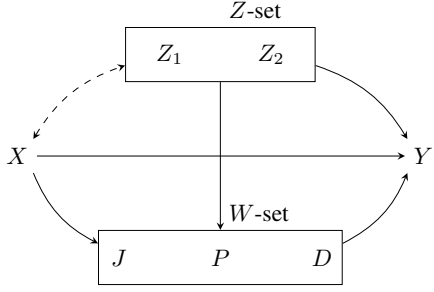
\begin{figure}
    \centering
    \scalebox{0.9}{
    \begin{tikzpicture}
	 	[>=stealth, rv/.style={thick}, rvc/.style={triangle, draw, thick, minimum size=8mm}, node distance=7mm]
	 	\pgfsetarrows{latex-latex};
	 	\begin{scope}
	 	\node[rv] (c) at (2.25,1.5) {${Z_1}$};
	 	\node[rv] (z2) at (3.75,1.5) {${Z_2}$};
	 	\node[rv] (a) at (0,0) {$X$};
	 	\node[rv] (m) at (1.5,-1.5) {$J$};
	 	\node[rv] (l) at (3,-1.5) {$P$};
	 	\node[rv] (r) at (4.5,-1.5) {${D}$};
	 	\node[rv] (y) at (6,0) {$Y$};
	 	
	 	\node (Zset) [draw,rectangle,minimum width=2.8cm,minimum height=0.8cm,label={[above right]$Z$-set}] at (3,1.5) {};
	 	\node (Wset) [draw,rectangle,minimum width=3.6cm,minimum height=0.8cm, label={[above right]$W$-set}] at (3,-1.5) {};
	 	
	 	\path[->] (a) edge[bend left = 0] (y);
	 	\path[->] (a) edge[bend left = -20] (Wset);
	 	\path[->] (Zset) edge[bend left = 0] (Wset);
	 	\path[->] (Wset) edge[bend left = -20] (y);
	 	\path[->] (Zset) edge[bend left = 20] (y);
	 	
	 	\path[<->,dashed] (a) edge[bend left = 20](Zset);
	 	\end{scope}
	\end{tikzpicture}
    }
    \caption{SFM for the COMPAS dataset.}
    \label{fig:compassfm}
\end{figure}

We construct the standard fairness model (SFM) for this example, which is shown in Fig.~\ref{fig:compassfm}. The bidirected arrow between $X$ and $\{Z_1, Z_2 \}$ indicates possible co-variations of race with age and sex, which may not be causal in nature\footnote{The causal model is non-committal regarding the complex historical/social processes that lead to such co-variations.}. Furthermore, $\{Z_1, Z_2 \}$ are the confounders, not causally affected by race $X$. The set of mediators $\{J, P, D\}$, however, may be affected by race, due to an existing societal bias in policing and criminal justice. Finally, all of the above-mentioned variables may influence the outcome $Y$.

Having access to data from Broward County, and equipped with Alg.~\ref{algo:sp-to-pp-bn}, we want to prove that the recidivism predictions produced by Northpointe (labeled $\widehat{Y}^{NP}$) violate legal doctrines of anti-discrimination. Suppose that in an initial hearing, the Broward County district court determines that the direct and indirect effects are not in the business necessity set, while the spurious effect is. In words, gender ($Z_1$) and age ($Z_2$) are allowed to be used to distinguish between the minority and majority groups when predicting recidivism, while other variables are not. If Northpointe's predictions are found to be discriminatory, we are required by the court to produce better, non-discriminatory predictions.

\begin{figure}
    \centering
    \includegraphics[width=0.9\columnwidth]{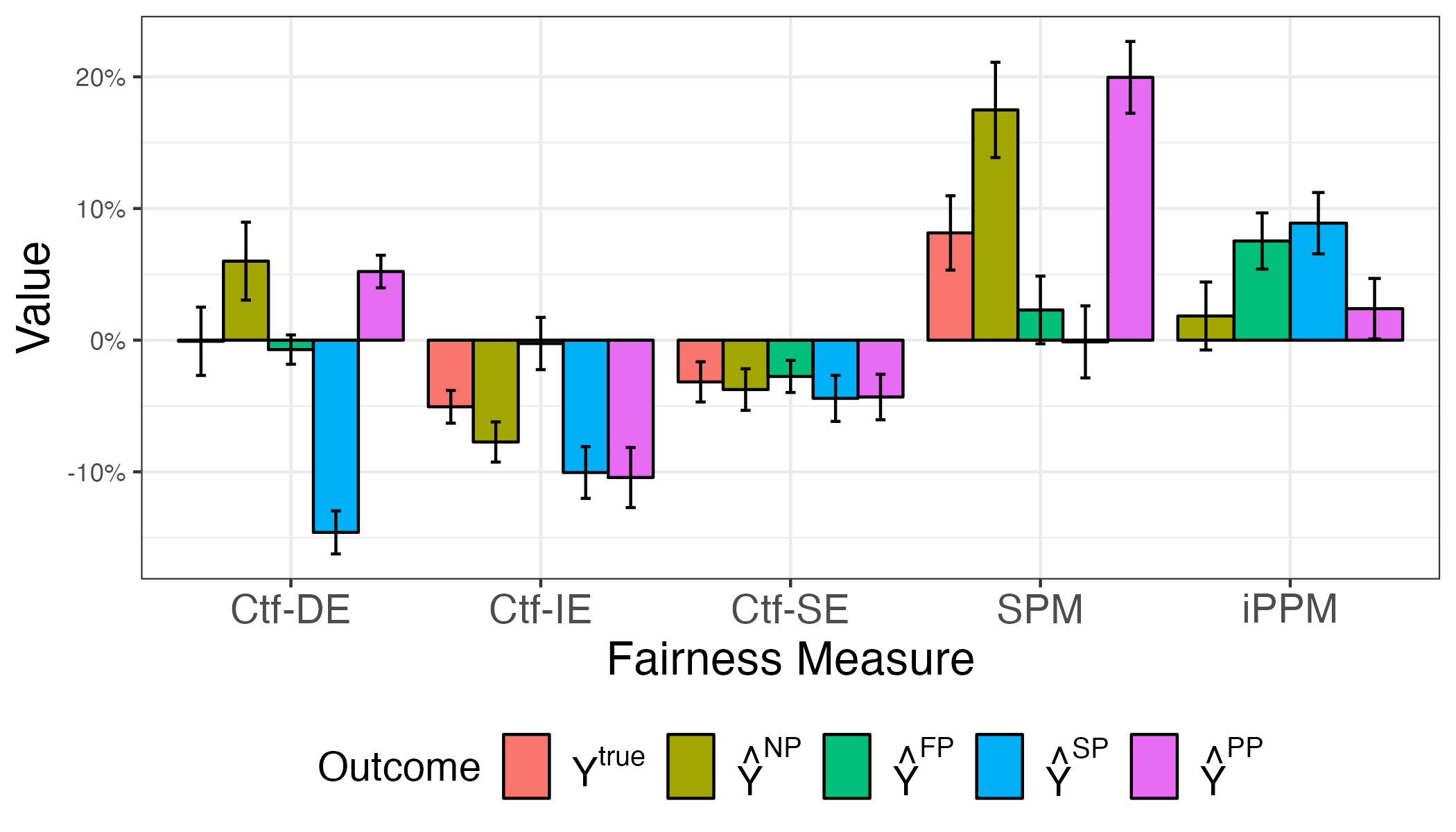}
    \caption{Fairness measures on COMPAS.}
    \label{fig:compas-effects-graph}
\end{figure}

In light of this information, we proceed as follows (see \elink{https://github.com/dplecko/sp-to-pp/blob/main/sp-pp-compas.R}{source code}). We first obtain a causally fair predictor $\widehat{Y}^{FP}$ using the \texttt{fairadapt} package, which performs optimal transport in order to remove the direct and indirect effects of the protected attribute from the prediction. Additionally, we also add predictors $\widehat{Y}^{SP}$ (obtained using group-specific thresholds), $\widehat{Y}^{PP}$ (the unconstrained predictor using random forests) that satisfy statistical and predictive parity, respectively. Then, we compute the counterfactual causal measures of fairness for the true outcome $Y$, Northpointe's predictions $\widehat{Y}^{NP}$, fair predictions $\widehat{Y}^{FP}$, and $\widehat{Y}^{SP}, \widehat{Y}^{PP}$ (see Fig.~\ref{fig:compas-effects-graph}). For the direct effect, we have:
\begin{align}
    \text{Ctf-DE}_{x_0, x_1}(y\mid x_0) &= -0.08 \% \pm 2.59 \%, \\
    \text{Ctf-DE}_{x_0, x_1}(\widehat{y}^{NP}\mid x_0) &= 6 \% \pm 2.96 \%, \\
    \text{Ctf-DE}_{x_0, x_1}(\widehat{y}^{FP}\mid x_0) &= -0.72 \% \pm 1.11 \%.
\end{align}
The indicated 95\% confidence intervals are computed using repeated bootstrap repetitions of the dataset. Since the direct effect is not in the business necessity set, Northpointe's predictions clearly violate the disparate treatment doctrine (yellow bar for the Ctf-DE measure in Fig.~\ref{fig:compas-effects-graph}). Our predictions, however, do not exhibit a statistically significant direct effect of race on the outcome, so they do not violate the criterion (green bar). Next, for the indirect effects, we obtain:
\begin{align}
    \text{Ctf-IE}_{x_1, x_0}(y\mid x_0) &= -5.06 \% \pm 1.24 \%, \\
    \text{Ctf-IE}_{x_1, x_0}(\widehat{y}^{NP}\mid x_0) &= -7.73 \% \pm 1.53 \%, \\
    \text{Ctf-IE}_{x_1, x_0}(\widehat{y}^{FP}\mid x_0) &= -0.25 \% \pm 1.98 \%.
\end{align}
Once again, the indirect effect, which is not in the business necessity set, is different from $0$ for the Northpointe's predictions (violating disparate impact, see yellow bar for Ctf-IE in Fig.~\ref{fig:compas-effects-graph}), but not statistically different from $0$ for our predictions (green bar). Interestingly, the indirect effect is different from $0$ for the true outcome (red bar), indicating a bias in the current real world. Finally, for spurious effects we obtain
\begin{align}
    \text{Ctf-SE}_{x_1, x_0}(y) &= -3.17 \% \pm 1.53 \%, \\
    \text{Ctf-SE}_{x_1, x_0}(\widehat{y}^{NP}) &= -3.75 \% \pm 1.58 \%, \\
    \text{Ctf-SE}_{x_1, x_0}(\widehat{y}^{FP}) &= -2.75 \% \pm 1.22 \%.
\end{align}
Since the spurious effect is in the business necessity set, and the hypotheses $\text{SE}(y) = \text{SE}(\widehat y^{NP})$, $\text{SE}(y) = \text{SE}(\widehat y^{FP})$ are not rejected (p-values $0.30, 0.32$, respectively), no violations with respect to spurious effects are found.

We conclude that Northpointe's predictions $\widehat{Y}^{NP}$ violate the legal doctrines of fairness, while our predictions $\widehat{Y}^{FP}$ do not. Importantly, tying back to the original discussion that motivated our approach, Northpointe's predictions $\widehat{Y}^{NP}$ are further away from statistical parity than our predictions $\widehat{Y}^{FP}$ according to the SPM (see SPM column in Fig.~\ref{fig:compas-effects-graph}), while at the same time better calibrated according to the integrated PP measure (iPPM) that averages the PP measures from Eq.~\ref{eq:pp-measure} across different values of $\widehat{y}$ (see iPPM column in the figure). This demonstrates the trade-off between statistical and predictive parity through business necessity.

The choice of the business necessity set in the example above, BN $ = \{ W \}$, was arbitrary. In general, the BN set can be any of $\emptyset, Z, W, \{Z, W \}$, which we explore next.
\paragraph{Statistical vs. Predictive Parity Pareto Frontier.} \label{appendix:pareto-frontier}
We now investigate other possible choices of the BN set, namely BN sets $\emptyset, Z, W, \{Z, W \}$. Based on the theoretical analysis of Sec.~\ref{sec:causal-decompositions}, we expect that different choices of the BN set will lead to different trade-offs between SP and PP.

In particular, with the BN set being empty, any variation from $X$ to $Y$ would be considered discriminatory, so one would expect to be close to statistical parity. In contrast, if the BN set is $\{Z, W\}$, then all variations (apart from the direct effect) are considered as non-discriminatory, so one would expect to be closer to satisfying predictive parity. The remaining options, BN$=W$, and BN$=Z$ should interpolate between these two extremes.

Based on the above intuition, we proceed as follows. For each choice of the business necessity set,
\begin{align}
    \text{BN} \in \{ \emptyset, Z, W, \{Z, W \} \},
\end{align}
we compute the adjusted version of the data again using the \texttt{fairadapt} package, with the causal effects in the complement of the BN set being removed. That is, if BN = $W$, then our procedure removes the spurious effect, but keeps the indirect effect intact (other choices of the BN set are similar). Therefore, for each BN set, we obtain an appropriately adjusted predictor $\widehat{Y}_{\text{BN}}^{FP}$, and in particular compute the predictors
  $ \widehat{Y}_{\emptyset}^{FP}, \widehat{Y}_{Z}^{FP}, \widehat{Y}_{W}^{FP}, \text{ and } \widehat{Y}_{\{Z, W\}}^{FP}. $
We note that the fair predictor $\widehat{Y}^{FP}$ from the first part of the example is the predictor with the BN set equal to $Z$, i.e., it corresponds to $\widehat{Y}_{Z}^{FP}$. For each $\widehat{Y}_{\text{BN}}^{FP}$ we compute the SPM, iPPM, and area under receiver operator characteristic (AUC):
\begin{align}
    \text{SPM}_{x_0, x_1}(\widehat{y}_{\text{BN}}^{FP}), \text{iPPM}_{x_0, x_1}(\widehat{y}_{\text{BN}}^{FP}), \text{AUC}(\widehat{y}_{\text{BN}}^{FP}),
\end{align}
across 10 different repetitions of the adjustment procedure that yields $\widehat{Y}_{\text{BN}}^{FP}$. For each business necessity set, this allows us to compute the SPM (measuring statistical parity), and iPPM (measuring predictive parity). 

\begin{figure}
    \centering
    \includegraphics[width=0.9\linewidth]{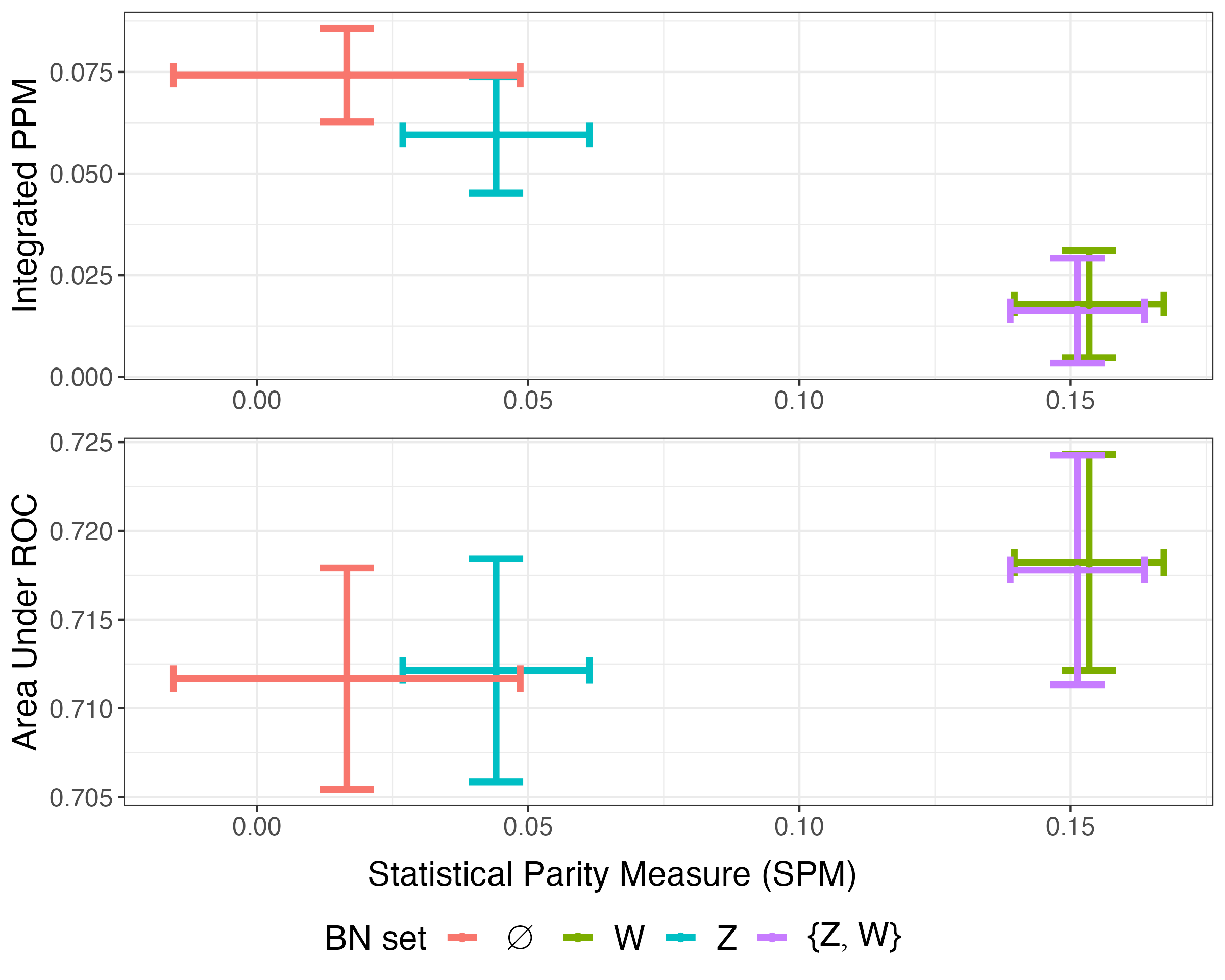}
    \caption{SP vs. PP and AUC Pareto frontiers on COMPAS.}
    \label{fig:pareto-frontier}
    \vspace{-0.15in}
\end{figure}
The results of the experiment are shown in Fig.~\ref{fig:pareto-frontier}, where the error bars indicate the standard deviation over different repetitions. As predicted by our theoretical analysis, the choice of BN$=\emptyset$ yields the lowest SPM, but the largest iPPM. Conversely, BN$=\{Z, W\}$ yields the lowest iPPM, but the largest SPM. The BN sets $Z, W$ interpolate between the two notions, but the data indicates the spurious effect explained by $Z$ does not have a major contribution. Fig.~\ref{fig:pareto-frontier}, therefore, shows a trade-off between statistical parity on the one hand, and predictive parity and accuracy on the other. The trade-off is captured through different business necessity options, and gives an empirical validation of the hypothesized spectrum of fairness notions in Fig.~\ref{fig:sp-pp-spectrum}.

\section{Conclusions} \label{sec:conclusions}
The literature in fair ML is abundant with fairness measures \citep{corbett2018measure}, many of which are mutually incompatible. Nonetheless, it is doubtful that each of these measures corresponds to a fundamentally different ethical conception of fairness. The multitude of possible approaches to quantifying discrimination makes the consensus on an appropriate notion of fairness unattainable. Further, the impossibility results between different measures may be discouraging to data scientists who wish to quantify and remove discrimination, but are immediately faced with a choice of which measure they wish to subscribe to. Our work falls under the rubric of analyzing impossibility results through a causal lens \citep{makar2022fairness}.

In particular, we focus on the impossibility of simultaneously achieving SP and PP. As our discussion shows, the guiding idea behind SP is that variations transmitted along causal pathways from the protected attribute to the predictor should equal $0$, i.e., the decision should not depend on the protected attribute through the causal pathway in question (Def.~\ref{def:causal-sp}). Complementary to this notion, and based on Thm.~\ref{thm:pp-decomposition}, the guiding principle behind PP is that variations transmitted along a causal pathway should be equal for the predictor as they are for the outcome \textit{in the real world} (Def.~\ref{def:causal-pp}). PP will therefore be satisfied when the BN set includes all variations coming from $X$ to $Y$, while SP will be satisfied when the BN set is empty. The choice of the BN set interpolates between SP and PP forming a spectrum of fairness notions (see Fig.~\ref{fig:sp-pp-spectrum}), in a way that can be formally assessed based on Alg.~\ref{algo:sp-to-pp-bn}. 

Therefore, our work complements the previous literature by reconciling the impossibility result between SP and PP \citep{barocas2017fairness}. Furthermore, it complements the existing literature on path-specific notions of fairness \citep{kilbertus2017avoiding, wu2019pc, chiappa2019path}, which does not consider the true outcome $Y$ and the predictor $\widehat{Y}$ simultaneously, and does not explicitly specify which levels of discrimination are deemed acceptable along causal pathways in the BN set, as described in Alg.~\ref{algo:sp-to-pp-bn}.

\bibliography{refs}
\paragraph{Acknowledgements} This research was supported in part by the NSF, ONR, AFOSR, DARPA, DoE, Amazon, JP Morgan, and The Alfred P. Sloan Foundation.

\paragraph{Supplements} The supplementary materials for the paper can be accessed in the \elink{https://arxiv.org/pdf/2306.05059.pdf}{arXiv version}.

\newpage
\appendix
\section*{\centering\Large Supplementary Material for \textit{Reconciling Predictive and Statistical Parity: A Causal Approach}}
The source code for reproducing all the experiments can be found in our \elink{https://github.com/dplecko/sp-to-pp}{code repository}. All the code was ran on MacOS 13 Ventura, with 16GB RAM and 2.2GHz 6-core Intel Core i7.
\section{Thm.~\ref{thm:pp-decomposition} Proof and Evaluation}

\subsection{Proof of Thm.~\ref{thm:pp-decomposition}} \label{appendix:pp-decomp-proof}
\begin{proof}
We prove the theorem and the corollary jointly. Note that term $\ex(y \mid x_1, \widehat{y}) - \ex(y \mid x_0, \widehat{y})$ equals
\begin{align}
    &= \underbrace{\ex(y_{x_1} \mid x_1, \widehat{y}_{x_1}) - \ex(y_{x_0} \mid x_1, \widehat{y}_{x_1})}_{\text{Term (I)}}\\
    &+ \underbrace{\ex(y_{x_0} \mid x_1, \widehat{y}_{x_1}) - \ex(y_{x_0} \mid x_1, \widehat{y}_{x_0})}_{\text{Term (II)}} \\
    &+ \underbrace{\ex(y_{x_0} \mid x_1, \widehat{y}_{x_0}) - \ex(y_{x_0} \mid x_0, \widehat{y}_{x_0})}_{\text{Term (III)}}.
\end{align}
Since there are no backdoor paths between $X$ and $Y, \widehat{Y}$, Term (III) vanishes.
By noting that $\ex(y_{x} \mid x_1, \widehat{y}_{x_1}) = \ex(y_{x} \mid x_1, \widehat{y})\;\;\forall x$ by consistency axiom \citep[Ch.7]{pearl:2k} (and applying it to Term (I)), and also that $Y_{x} \ci X$ (and applying it to Term (II)) gives us the theorem. The counterfactual independence $Y_x \ci X$ can be verified from twin-network \citep{pearl:2k}:
\begin{center}
    \begin{tikzpicture}
	 [>=stealth, rv/.style={thick}, rvc/.style={triangle, draw, thick, minimum size=7mm}, node distance=18mm]
	 \pgfsetarrows{latex-latex};
	 \begin{scope}
	 	\node[rv] (x) at (-1.5,0) {$X$};
	 	\node[rv] (w) at (0,0) {$W$};
	 	\node[rv] (y) at (1.5,0) {$Y$};

            \node[rv] (wx) at (0,1.5) {$W_x$};
	 	\node[rv] (yx) at (1.5,1.5) {$Y_x$};
            
	 	\draw[<->, dashed] (wx) -- (w);
            \draw[<->, dashed] (yx) -- (y);
            \draw[->] (x) -- (w);
	 	\draw[->] (x) edge[bend right = 20] (y);
		\draw[->] (w) -- (y);
	 \end{scope}
	 \end{tikzpicture}
\end{center}
from which we verify that $Y_x$ and $X$ are d-separated by the empty set $\emptyset$, implying $Y_x \ci X$.

For the corollary, we further assume that the SCM is linear, and that the predictor $\widehat{Y}$ is efficient, i.e., $\widehat{Y}(x, w) = \ex[Y \mid x, w]$. In this case, the efficiency simply translates to the fact that
\begin{align}
    \alpha_{W\widehat{Y}} = \alpha_{WY}, \alpha_{X\widehat{Y}} = \alpha_{XY}.
\end{align}
Due to linearity, for every unit $u$, we have that
\begin{align}
    y_{x_1}(u) - y_{x_0}(u) = \alpha_{XW}\alpha_{WY} + \alpha_{XY},
\end{align}
and since Term (I) can be written as $\sum_{u} [y_{x_1}(u) - y_{x_0}(u)]P(u \mid x_1, \widehat{y})$ using the unit-level expansion of counterfactual distributions \citep{tian2000probabilities, bareinboim2020on}, Eq.~\ref{eq:pp-causal} follows. Similarly, Term (II) can be expanded as
\begin{align}
    \sum_u \widehat{y}_{x_0}(u) [P(u \mid \widehat{y}_{x_1}) - P(u \mid\widehat{y}_{x_0})].
\end{align}
We now look at units $u$ which are compatible with $\widehat{Y}_{x_1}(u) = \widehat{y}$ and $\widehat{Y}_{x_0}(u) = \widehat{y}$. We can expand $\widehat{Y}_{x_1}(u)$ as
\begin{align}
    \widehat{Y}_{x_1}(u) = \alpha_{X\widehat{Y}} + \alpha_{XW}\alpha_{W\widehat{Y}} + \alpha_{W\widehat{Y}}u_W.
\end{align}
Thus, we have that
\begin{align}
    \widehat{Y}_{x_1}(u) = \widehat{y} \implies \alpha_{W\widehat{Y}}u_W = \widehat{y} - \alpha_{X\widehat{Y}} + \alpha_{XW}\alpha_{W\widehat{Y}}. \label{eq:yx1-hat}
\end{align}
Similarly, we also obtain that
\begin{align}
    \widehat{Y}_{x_0}(u) = \widehat{y} \implies \alpha_{W\widehat{Y}}u_W = \widehat{y}. \label{eq:yx0-hat}
\end{align}
Due to the efficiency of learning which implies that $\alpha_{W\widehat{Y}} = \alpha_{WY}$ and $\alpha_{X\widehat{Y}} = \alpha_{XY}$, Eq.~\ref{eq:yx1-hat} and \ref{eq:yx0-hat} imply
\begin{align}
    y_{x_0}(u) &= \widehat{y} - (\alpha_{XY} + \alpha_{XW}\alpha_{WY})\;\; \forall u \text{ s.t. } \widehat{Y}_{x_1}(u) = \widehat{y}, \\
    y_{x_0}(u) &= \widehat{y} \;\; \forall u \text{ s.t. } \widehat{Y}_{x_0}(u) = \widehat{y},
\end{align}
which in turn shows that
\begin{align}
    \ex(y_{x_0} \mid \widehat{y}_{x_1}) - \ex(y_{x_0} \mid \widehat{y}_{x_0}) = -\alpha_{XY} - \alpha_{XW}\alpha_{WY}.
\end{align}
\end{proof}
\subsection{Empirical Evaluation of Thm.~\ref{thm:pp-decomposition}} \label{appendix:pp-decomp-empirical}
In this section, we empirically demonstrate the validity of Thm.~\ref{thm:pp-decomposition} through an example (see \elink{https://github.com/dplecko/sp-to-pp/blob/main/pp-theorem-empirical.R}{source code} for reproducing the results). In particular, we consider the following structural causal model:
\begin{empheq}[left ={\mathcal{F}^*, P^*(U): \empheqlbrace}]{align}
		        X  &\gets U_X \label{eq:synth-first} \\
                    W_1 &\gets X + U_1  \\
 		        W_2 & \gets \frac{1}{4}W_1^2 - \frac{1}{3}X + U_2 \\
 		        Y  &\gets \frac{1}{6}W_1W_2 + W_1 + \frac{1}{2}X + U_Y. \\
 		            U_X &\in \{0,1\}, P(U_X = 1) = 0.5, \\
                        U_1&, U_2, U_Y \sim N(0, 1),  \label{eq:synth-last}
\end{empheq}
with the following causal diagram:
\begin{center}
    \begin{tikzpicture}
	 [>=stealth, rv/.style={thick}, rvc/.style={triangle, draw, thick, minimum size=7mm}, node distance=18mm]
	 \pgfsetarrows{latex-latex};
	 \begin{scope}
	 	\node[rv] (x) at (-1.5,-1) {$X$};
	 	\node[rv] (w1) at (-0.6,-2) {$W_1$};
            \node[rv] (w2) at (0.6,-2) {$W_2$};
	 	\node[rv] (y) at (1.5,-1) {$Y$};
	 	\draw[->] (x) -- (w1);
            \draw[->] (x) -- (w2);
            \draw[->] (w1) -- (w2);
	 	\draw[->] (x) -- (y);
		\draw[->] (w1) -- (y);
            \draw[->] (w2) -- (y);
	 \end{scope}
	 \end{tikzpicture}.
\end{center}
We proceed as follows. We investigate the decomposition in Thm.~\ref{thm:pp-decomposition} with respect to a varying number of samples available, with $n \in \{ 1000, 2000, 3500, 5000, 7500, 10000 \}$. For each value, we generate $n$ samples from the SCM in Eqs.~\ref{eq:synth-first}-\ref{eq:synth-last}. Then, based on the samples, we construct a predictor $\widehat{Y}$ using \texttt{xgboost} \citep{chen2016xgboost} (with a fixed $\eta = 0.1$ and number of rounds chosen using cross-validation with 10 folds). Then, we bin the samples into 20 intervals corresponding to the quantiles of the predictor $\widehat{Y}$:
\begin{align*}
    b_1 = [0\%\text{-quant}(\widehat{Y})&, 5\%\text{-quant}(\widehat{Y})), \\ 
    &\vdots  \\
    b_{20} = [95\%\text{-quant}(\widehat{Y})&, 100\%\text{-quant}(\widehat{Y})].
\end{align*}
Further, we estimate the potential outcome $Y_{x_0}$ for each sample using the \texttt{fairadapt} package \citep{plevcko2021fairadapt}, and the estimator is labeled $Y^*_{x_0}$ (also note that $Y_{x_0}(u) = Y(u)$ for all $u$ s.t. $X(u) = x_0$). Then, within each bin $b_i$, we compute
\begin{figure}
    \centering
    \includegraphics[width=0.95\columnwidth]{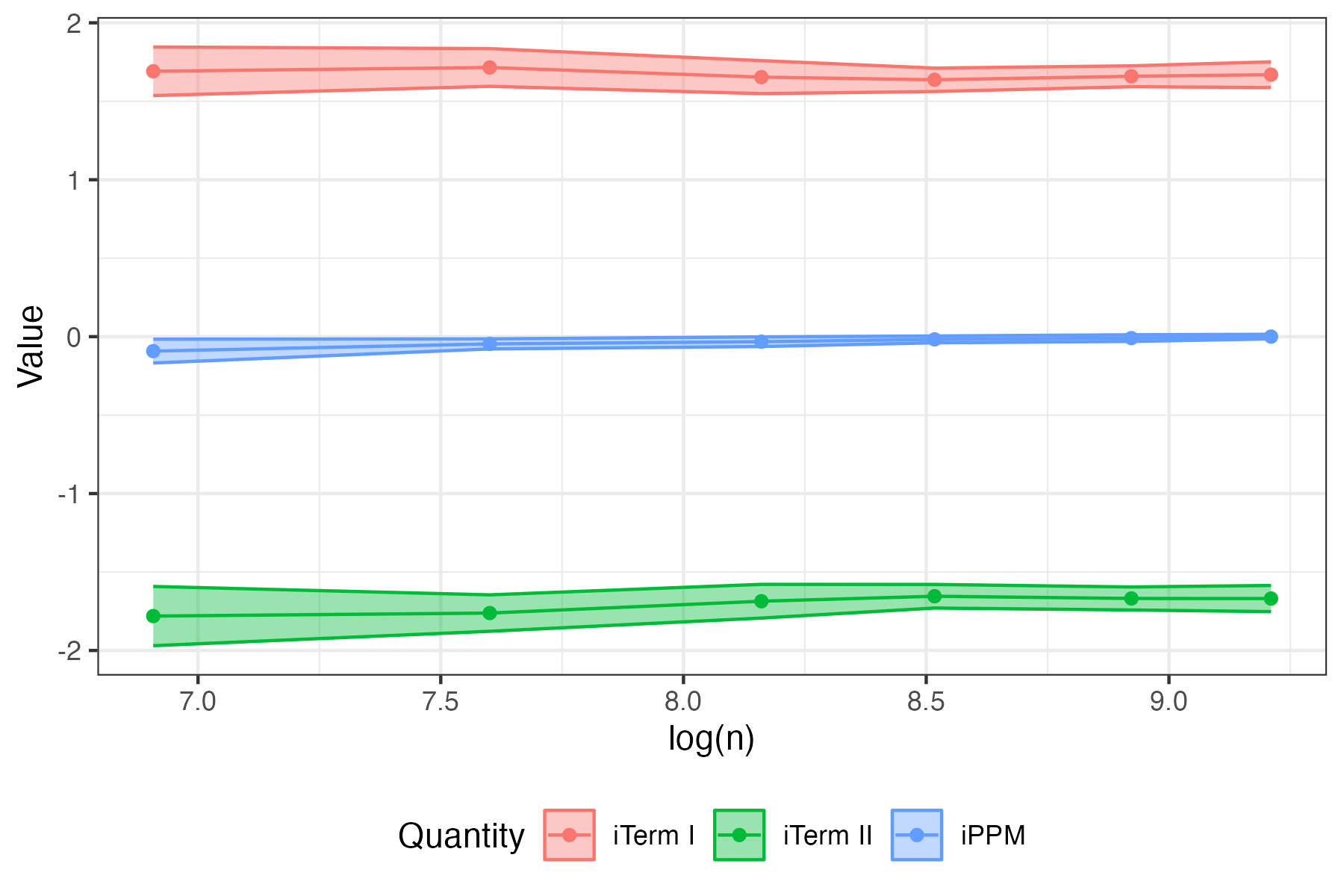}
    \caption{iTerm I, iTerm II, iPPM estimated values.}
    \label{fig:pp-decomp-fig1}
\end{figure}
\begin{align}
    \widehat{\text{Term I}}_i &= \widehat{\ex} [Y_{x_1} \mid x_1, \widehat{y} \in b_i] - \widehat{\ex} [Y^*_{x_0} \mid x_1, \widehat{y} \in b_i], \label{eq:termI-hat} \\
    \widehat{\text{Term II}}_i &= \widehat{\ex} [Y^*_{x_0} \mid x_1, \widehat{y} \in b_i] - \widehat{\ex} [Y_{x_0} \mid x_0, \widehat{y} \in b_i],\\
    \widehat{\text{iPPM}}_i &= \widehat{\ex} [Y_{x_1} \mid x_1, \widehat{y} \in b_i] - \widehat{\ex} [Y_{x_0} \mid x_0, \widehat{y} \in b_i],
\end{align}
where $\widehat{\ex}$ denotes the empirical expectation.
We then compute $\widehat{\text{iTerm I}} = \frac{1}{20} \sum_{i=1}^{20} \widehat{\text{Term I}}_i$, where iTerm I is the ``integrated'' value of Term I across all 20 bins. We similarly estimate $\widehat{\text{iTerm II}}$ and $\widehat{\text{iPPM}}$.
For each dataset size $n$, the experiment is repeated 20 times. In Fig.~\ref{fig:pp-decomp-fig1}, we plot the obtained estimates of iTerm I, iTerm II, and the iPPM, where the shaded region represents the standard deviation of the estimates across 20 repetitions. The figure illustrates that as the number of samples increases, the predictor $\widehat{Y}$ is closer to calibration (iPPM is closer to $0$), as predicted by Prop.~\ref{prop:pp-efficient}. Furthermore, we can see that iTerm I and iTerm II are different from $0$, and cancel each other out, as predicted by Thm.~\ref{thm:pp-decomposition}.

\begin{figure}
   \centering
    \includegraphics[width=0.95\columnwidth]{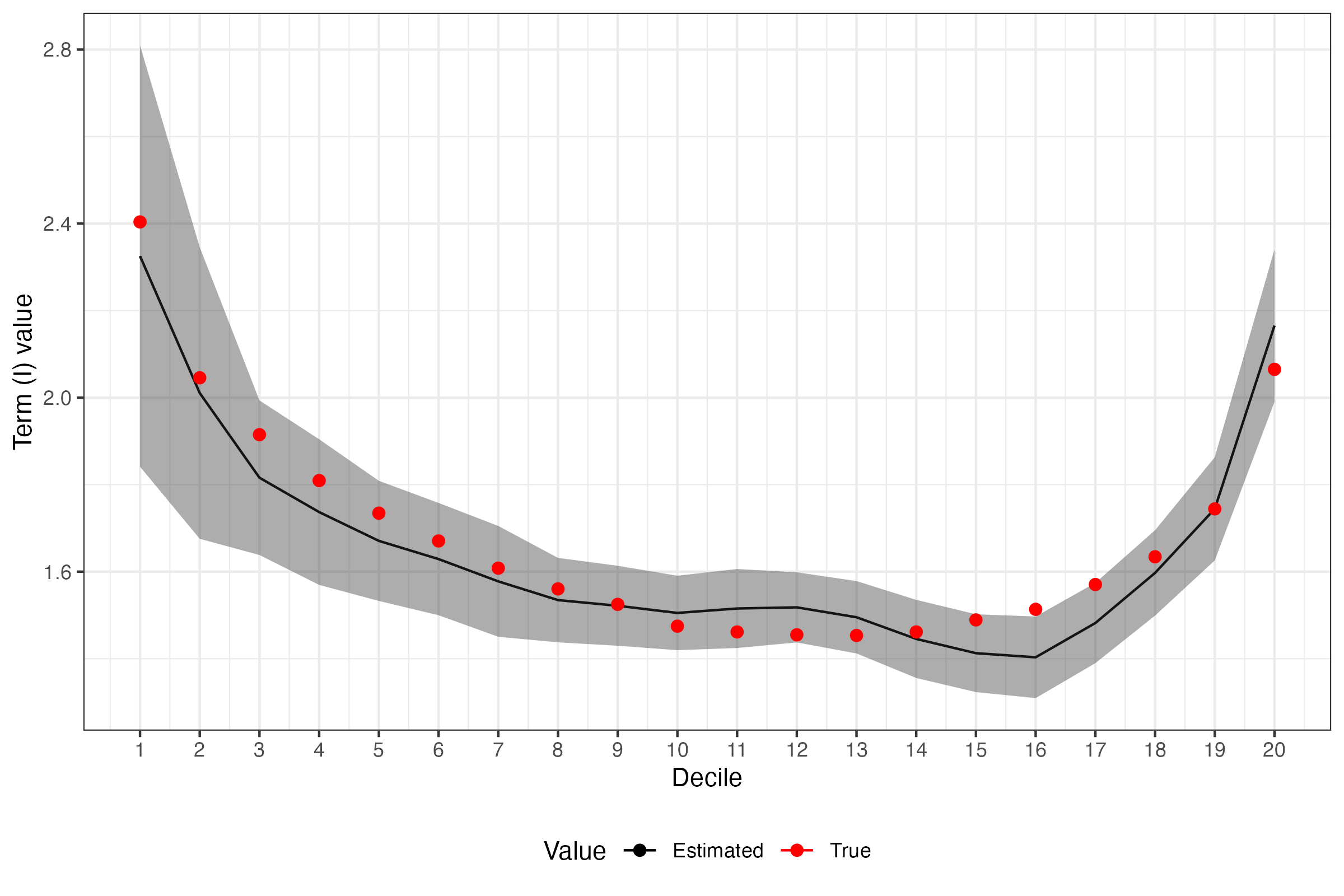}
    \caption{iTerm I and $\widehat{\text{Term I}}_i$ values.}
    \label{fig:pp-decomp-fig2}
\end{figure}

Furthermore, we verify if the estimates $\widehat{\text{Term I}}_i$ obtained in Eq.~\ref{eq:termI-hat} correspond to the ground truth. To this end, instead of estimating the potential outcome $Y_{x_0}$, we obtain the true potential outcome values $Y_{x_0}$ from the SCM. Then, we compute
\begin{align}
     \text{Term I}_i &= \ex [Y_{x_1} \mid x_1, \widehat{y} \in b_i] - \ex [Y_{x_0} \mid x_1, \widehat{y} \in b_i] \label{eq:termI}.
\end{align}
We take 50 repetitions of $n = 5000$ samples and obtain the values of $\widehat{\text{Term I}}_i$ and $\text{Term I}_i$. The obtained results are shown in Fig.~\ref{fig:pp-decomp-fig2}. In red, we can see the ground truth values for each $\text{Term I}_i$, and in black we see the estimated values of $\widehat{\text{Term I}}_i$, where the shaded area indicates the standard deviation of the estimate across the 50 repetitions. Once again, we see that the terms $\text{Term I}_i$ are estimated correctly, which in combination with Fig.~\ref{fig:pp-decomp-fig1} demonstrates empirically the theoretical result predicted by Thm.~\ref{thm:pp-decomposition}.
\section{Identification Expressions of Prop.~\ref{prop:ctf-family-id}} \label{appendix:measure-id-expressions}
We provide the identification expressions for the counterfactual measures $\text{Ctf-DE}_{x_0, x_1}(y\mid x)$, $\text{Ctf-IE}_{x_1, x_0}(y\mid x)$, and $\text{Ctf-SE}_{x_1, x_0}(y)$. For the measures related measures of $\widehat{y}$, the expressions are analogous, with $y$ replaced by $\widehat{y}$ throughout:
\begin{align}
        \text{Ctf-DE}_{x_0, x_1}(y\mid x) =& \sum_{z, w} [P(y \mid x_1, z, w) - P(y\mid x_0, z, w)] \nonumber \\
        & \quad \cdot P(w \mid x_0, z)P(z \mid x), \label{eq:ctf-de-id} \\
			\text{Ctf-IE}_{x_1, x_0}(y\mid x) =& \sum_{z, w} P(y\mid x_1, z, w) \nonumber\\
                &\quad \cdot [P(w \mid x_0, z) - P(w \mid x_1, z)] \nonumber\\ 
                & \quad \cdot P(z \mid x), \label{eq:ctf-ie-id}\\
            \text{Ctf-SE}_{x_1, x_0}(y) =& \sum_z P(y \mid x_1, z)[P(z \mid x_1) - P(z \mid x_0)] \label{eq:ctf-se-id}.
\end{align}
\section{Proof of Prop.~\ref{prop:pp-efficient}} \label{appendix:prop1-proof}
\begin{proof}
Notice that for any $X = x$
\begin{align}
    P(y \mid x, \widehat{y}) &= \sum_{\substack{z, w: \\\widehat{Y}(x,z,w)=\widehat{y}}} P(y \mid x, z, w, \widehat{y})P(z, w \mid x, \widehat{y}) \label{eq:step-1} \\ 
    &= \sum_{\substack{z, w: \\\widehat{Y}(x,z,w)=\widehat{y}}}P(y \mid x, z, w)P(z, w \mid x, \widehat{y}) \label{eq:step-2} \\
    &= \widehat{y} * \sum_{\substack{z, w: \\\widehat{Y}(x,z,w)=\widehat{y}}} P(z, w \mid x, \widehat{y}) = \widehat{y}. \label{eq:step-3}
\end{align}
The first step (Eq.~\ref{eq:step-1}) follows from the law of total probability, the second (Eq.~\ref{eq:step-2}) from noting that $\widehat{Y} \ci Y \mid X,Z,W$, and the third (Eq.~\ref{eq:step-3}) from the efficiency of the learner (Eq.~\ref{eq:efficient-learner}). Therefore, it follows that $P(y \mid x_1, \widehat{y}) = P(y \mid x_0, \widehat{y})$, meaning that $\widehat{Y}$ satisfies PP. 
\end{proof}
\section{Binary vs. Continuous $\widehat{Y}$} \label{appendix:binary-vs-cts}
In this appendix we discuss how the main results of the paper depend on whether the predictor $\widehat{Y}$ is continuous (such as in regression or probability prediction) or binary (such as in classification). For instance, the conditional independence statement
\begin{align}
    Y \ci X \mid \widehat Y,
\end{align}
usually called \textit{sufficiency} \citep{barocas2017fairness} has different implications if $\widehat{Y}$ is continuous vs. binary. In particular, the key difference lies in the number of constraints implied by the sufficiency independence statement. The statement can be re-written as
\begin{align}
    P(y \mid x_1, \widehat{y}) = P(y \mid x_0, \widehat{y}) \; \forall \widehat{y}
\end{align}
meaning that the number of constraints is equal to the number of levels of $\widehat y$ due to the quantifier $\forall \widehat{y}$. In other words, a classifier would have two levels $\hat{y} = 0$ and $\hat{y} = 1$, whereas for a score-based predictor we would have $\widehat{y} \in [0,1]$ (or a number of bins in practice). We now discuss, result by result, on what the implications are for the binary vs. continuous distinction:
\begin{enumerate}[label=(\arabic*)]
    \item Prop.~\ref{prop:pp-efficient}: The Bayes optimal $P(y \mid x,z,w)$ predictor needs to be continuous for this statement, as mentioned in the main text.
    \item Prop.~\ref{prop:dp-pp-impossibility}: Note that the statement depends only on the conditional independence statement $Y \perp\!\!\!\perp X \mid \widehat{Y}$; therefore, the result is valid for both continuous and binary settings.
    \item Prop.~\ref{prop:tv-decomp}: The decomposition can be used for both the binary and the continuous case, exchanging the corresponding sums/integrals accordingly.
    \item Thm.~\ref{thm:pp-decomposition}: The proof of the theorem is agnostic w.r.t. the distinction of a score-based or thresholded classifier; the difference, though, arises from the number of levels of $\widehat{y}$ that is considered.
    \item Def.~\ref{def:causal-pp}, Alg.~\ref{algo:sp-to-pp-bn}: The notion of causal predictive parity, and the Business Necessity Cookbook described in Sec.~\ref{sec:combining-sp-pp} are also unaffected by this distinction. Suppose even that we have $Y$ as a binary label, and have $\widehat{Y}$ as a continuous score-based prediction. Then, the equalities of the form $$\ex[y_{C_1} \mid E] - \ex[y_{C_0} \mid E] = \ex[\widehat{y}_{C_1} \mid E] - \ex[\widehat{y}_{C_0} \mid E]$$ as considered in Alg.~\ref{algo:sp-to-pp-bn} would still make sense, and would capture the essence of the implied fairness condition.
\end{enumerate}
\section{Existence of Fair Predictor for Alg.~\ref{algo:sp-to-pp-bn}} \label{appendix:algo-1-existence}
In this appendix, we show that a predictor $\widehat{Y}$ satisfying Def.~\ref{def:causal-pp} and Alg.~\ref{algo:sp-to-pp-bn} exists. In particular, we consider the setting in which the BN-set = \{DE, IE, SE\}, but relaxations of this requirement follow naturally. Before stating the main result, we introduce some notation. In particular, let $x\text{-DE}^{\text{ID}}_{x_0, x_1}(y \mid x_0), x\text{-IE}^{\text{ID}}_{x_1, x_0}(y \mid x_0)$, and $x\text{-SE}^{\text{ID}}_{x_1, x_0}(y)$ be the identification expressions of the effects given in Eqs.~\ref{eq:ctf-de-id}, \ref{eq:ctf-ie-id}, and \ref{eq:ctf-se-id}, respectively. Let $x\text{-DE}^{\text{ID}}_{x_0, x_1}(\widehat{y} \mid x_0), x\text{-IE}^{\text{ID}}_{x_1, x_0}(\widehat{y} \mid x_0)$, and $x\text{-SE}^{\text{ID}}_{x_1, x_0}(\widehat{y})$ be the analogous identification expressions for the $\widehat{y}$, which are obtained simply by replacing $y$ with $\widehat{y}$.
\begin{theorem}[Existence of $\widehat Y$ satisfying Alg.~\ref{algo:sp-to-pp-bn}] \label{thm:algo-1-exists}
Let $\mathcal{M}$ be an SCM compatible with the SFM. 
Let $\widehat{Y}$ be constructed as the optimal solution to
\begin{alignat}{2}
    \label{eq:inproc-causal-objective}
    \widehat{Y} = &\argmin_{f}        &\quad& \ex[Y - f(X, Z, W)]^2\\
    &\;\;\text{ s.t.} &      & \negquad\negquad x\text{-DE}^{\text{ID}}_{x_0, x_1}(\widehat{y} \mid x_0) = x\text{-DE}^{\text{ID}}_{x_0, x_1}(y \mid x_0) \\
    &&      & \negquad\negquad x\text{-IE}^{\text{ID}}_{x_1, x_0}(\widehat{y} \mid x_0) = x\text{-IE}^{\text{ID}}_{x_1, x_0}(y \mid x_0) \\
    &&      & \negquad\negquad x\text{-SE}^{\text{ID}}_{x_1, x_0}(\widehat{y}) = x\text{-SE}^{\text{ID}}_{x_1, x_0}(y),
\end{alignat}
where $x\text{-DE}^{\text{ID}}$, $x\text{-IE}^{\text{ID}}$, and $x\text{-SE}^{\text{ID}}$ represent the identification expressions of the corresponding measures. Then, the predictor $\widehat{Y}$ satisfies 
\begin{align} \label{eq:x-spec-cond-rep-2}
    \text{Ctf-DE}_{x_0, x_1}(\widehat{y} \mid x_0) =& \text{Ctf-DE}_{x_0, x_1}(y \mid x_0) \\
    \text{Ctf-IE}_{x_1, x_0}(\widehat{y} \mid x_0) =& \text{Ctf-IE}_{x_1, x_0}(y \mid x_0) \\
    \text{Ctf-SE}_{x_1, x_0}(\widehat{y}) =& \text{Ctf-SE}_{x_1, x_0}(y),
\end{align}
implying that $\widehat Y$ satisfies the conditions of Alg.~\ref{algo:sp-to-pp-bn} with the BN-set equal to \{DE, IE, SE\}.
\end{theorem}
\begin{proof}
    Note that $\widehat{Y}$ is a function of $X, Z, W$. Therefore, we can add $\widehat{Y}$ to the SFM, as in Fig.~\ref{fig:sfm}. Crucially, because $\widehat{Y}$ satisfies these assumptions, the value of the measure $\text{Ctf-DE}_{x_0, x_1}(\widehat{y} \mid x_0)$ is equal to the identification expression
    \begin{align}
        \text{Ctf-DE}_{x_0, x_1}(\widehat{y} \mid x_0) = x\text{-DE}^{\text{ID}}_{x_0, x_1}(\widehat{y} \mid x_0).
    \end{align}
    The analogous holds for indirect, and spurious effects. Furthermore, note that from the optimization procedure for $\widehat Y$ we also have that
    \begin{align}
        x\text{-DE}^{\text{ID}}_{x_0, x_1}(\widehat{y} \mid x_0) = x\text{-DE}^{\text{ID}}_{x_0, x_1}(y \mid x_0).
    \end{align}
    Finally, since $Y$ also satisfies the identification assumptions as in Prop.~\ref{prop:ctf-family-id}, it follows that
    \begin{align}
        x\text{-DE}^{\text{ID}}_{x_0, x_1}(y \mid x_0) = \text{Ctf-DE}_{x_0, x_1}(y \mid x_0),
    \end{align}
    in turn showing that
    \begin{align}
        \text{Ctf-DE}_{x_0, x_1}(\widehat{y} \mid x_0) = \text{Ctf-DE}_{x_0, x_1}(y \mid x_0).
    \end{align}
    The claim for indirect and spurious effects works analogously. 
\end{proof}
\par \noindent The adaptation of the statement in Thm.~\ref{thm:algo-1-exists} is possible for other BN sets. For instance, if the IE effect was not in the BN set, then would replace
\begin{align}
       x\text{-IE}^{\text{ID}}_{x_1, x_0}(\widehat{y} \mid x_0) = x\text{-IE}^{\text{ID}}_{x_1, x_0}(y \mid x_0) 
\end{align}
with the constraint
\begin{align}
    x\text{-IE}^{\text{ID}}_{x_1, x_0}(\widehat{y} \mid x_0) = 0.
\end{align}
This type of optimization would guarantee that the indirect effect of $X$ on $\widehat Y$ is $0$, instead of having it equal to the indirect effect of $X$ on $Y$ as in Thm.~\ref{thm:algo-1-exists}.
\section{Experiment on Census dataset} \label{appendix:census}
In this appendix, we perform an experiment on the Census dataset \citep{plevcko2021fairadapt}, similar to the experiment for the COMPAS dataset in Sec.~\ref{sec:experiment}. The American Community Survey of 2018 \citep{uscb2018acs}, carried out by the United States Census Beaurau, collected broad information about the US Government employees, including demographic information $Z$ ($Z_1$ for age, $Z_2$ for race, $Z_3$ for nationality), gender $X$ ($x_0$ female, $x_1$ male), marital and family status $M$, education information $L$, and work-related information $R$. 
The outcome $Y$ is a 0/1 indicator for whether the person is paid less or more than the dataset median of 42,000\$/year.
We begin by constructing the Standard Fairness Model (SFM) for this dataset, which is given by
\begin{equation}
    \langle X = \lbrace X \rbrace,  Z = \lbrace Z_1, Z_2, Z_3 \rbrace, W = \lbrace M, L, R\rbrace, Y = \lbrace Y \rbrace\rangle.
\end{equation}
In words, the set of confounders $Z$ includes age, race, and nationality, while the set of mediators $W$ includes family status, education, and work-related information. 
We then construct a predicor $\widehat{Y}$ of $Y$, using \texttt{xgboost} \citep{chen2016xgboost}, with a fixed $\eta = 0.1$ and the number of rounds chosen using 10-fold cross-validation. Additionally, a fair predictor $\widehat{Y}^{FP}$ is constructed using \texttt{fairadapt} \citep{plevcko2021fairadapt}, which ensures that direct and indirect effects of the protected attribute on the predictor are removed. We consider the confounders $Z$ to be in the business necessity set.

\begin{figure}
    \centering
    \includegraphics[width=\linewidth]{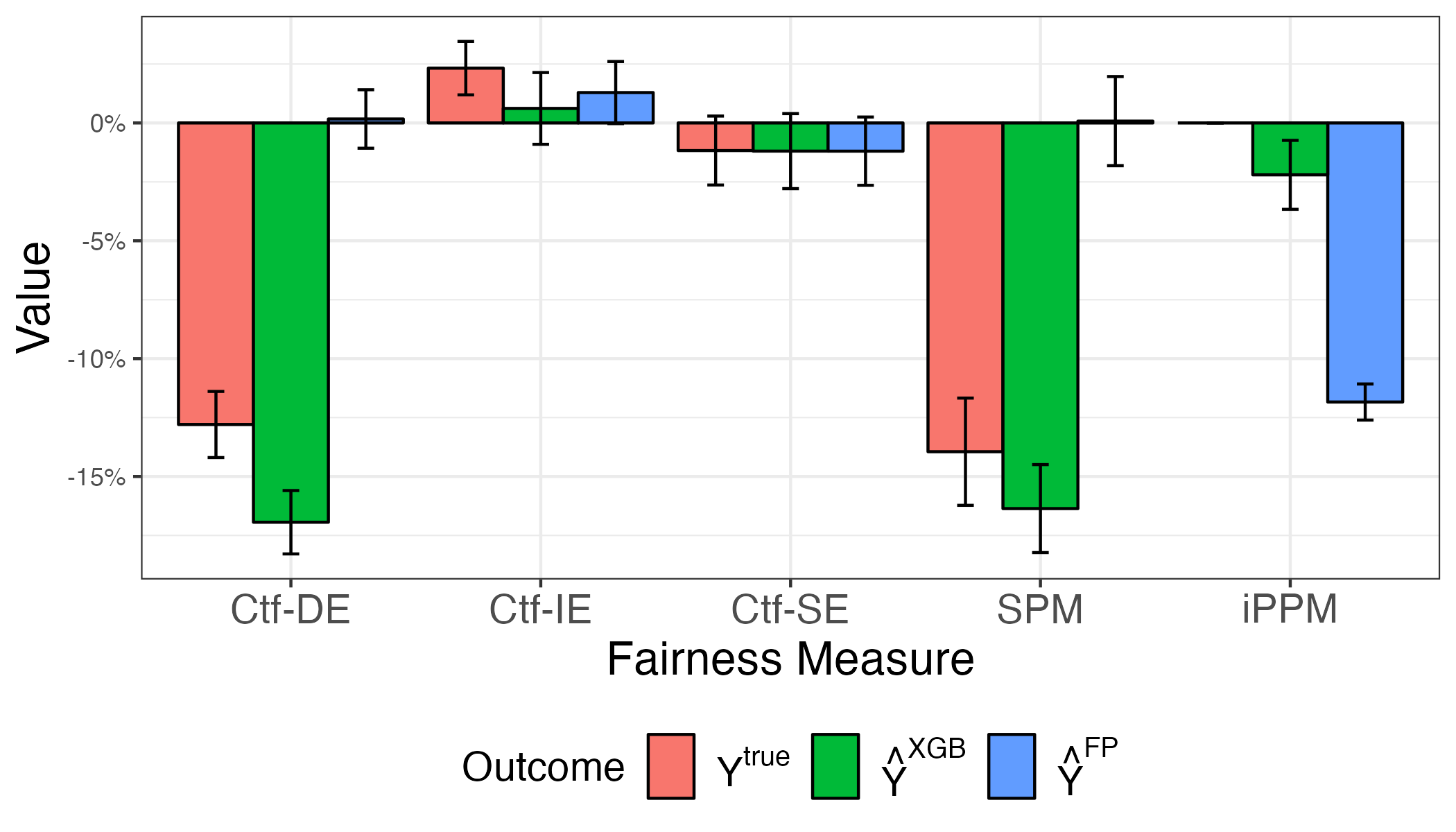}
    \caption{Alg.~\ref{algo:sp-to-pp-bn} applied to Census 2018 dataset.}
    \label{fig:census-algo-1}
\end{figure}

We then apply Alg.~\ref{algo:sp-to-pp-bn} and perform the causal decompositions of the SPM$_{x_0, x_1}(y)$, SPM$_{x_0, x_1}(\widehat y)$, and SPM$_{x_0, x_1}(\widehat y^{FP})$ measures into their direct, indirect, and spurious components. We also compute the iPPM measure for $Y, \widehat Y$, and $\widehat Y^{FP}$. The results are shown in Fig.~\ref{fig:census-algo-1}.
In particular, we can observe that the unconstrained predictor $\widehat{Y}$ has a large direct effect favoring men, same as the true outcome $Y$. $\widehat Y$ thus violates disparate treatment. At the same time, the fair predictor $\widehat Y^{FP}$ has no direct effect, and therefore does not violate disparate treatment.

For the indirect effects, we observe that the true indirect effect is different from $0$, while the indirect effects of $X$ on $\widehat Y$, $\widehat Y^{FP}$ are both not statistically different from $0$. Therefore, neither $\widehat Y$ nor $\widehat Y^{FP}$ violate disparate impact with respect to the indirect effects.

For the spurious effects, neither the outcome $Y$ nor its predictors $\widehat Y$, $\widehat Y^{FP}$ exhibit discrimination, i.e., none of the effects are statistically different from $0$. Therefore, there are no violations of disparate impact with respect to spurious effects.

Finally, we observe that the unconstrained predictor $\widehat Y$ has a low iPPM measure (i.e., it is close to predictive parity) while having a large SPM (further from statistical parity). In contrast to this, the fair predictor $\widehat Y^{FP}$ has a large iPPM measure (i.e., it is further away from predictive parity) while having a small SPM (closer to statistical parity). Once again, this demonstrates the spectrum between SP and PP discussed in the main text, which is spanned by different choices of the BN set.



\end{document}